\documentclass[twocolumn,showpacs,amsmath,amssymb,pra]{revtex4}
\usepackage{amsxtra}
\usepackage{amssymb}
\usepackage{amsmath}
\usepackage{graphicx}

\newcommand {\rhovec}{\ensuremath \boldsymbol{\rho}}
\newcommand {\kvec}{\ensuremath \boldsymbol{k}}

\begin{document}
\title{Unified Theory of Ghost Imaging with Gaussian-State Light}
\date{\today}
\author{Baris I. Erkmen}
\email{erkmen@mit.edu}
\author{Jeffrey H. Shapiro}
\affiliation{Massachusetts Institute of Technology, Research Laboratory of Electronics, Cambridge, Massachusetts 02139, USA}

\begin{abstract}
The theory of ghost imaging is developed in a Gaussian-state framework that both encompasses prior work---on thermal-state and biphoton-state imagers---and provides a complete understanding of the boundary between classical and quantum behavior in such systems.  The core of this analysis is the expression derived for the photocurrent-correlation image obtained using a general Gaussian-state source.  This image is expressed in terms of the phase-insensitive and phase-sensitive cross-correlations between the two detected fields, plus a background. Because any pair of cross-correlations is obtainable with classical Gaussian states, the image does not carry a quantum signature {\em per se}.  However, if the image characteristics of classical and nonclassical Gaussian-state sources with identical auto-correlation functions are compared, the nonclassical source provides resolution improvement in its near field and field-of-view improvement in its far field.
\end{abstract}
\maketitle 

\section{Introduction}

Ghost imaging is the acquisition of an object's transverse transmittance pattern by means of photocurrent correlation measurements. In a generic ghost imaging experiment (see the example in Fig.~\ref{GI:propagation}), a classical or quantum source that generates two paraxial optical fields is utilized. These fields propagate in two different directions, through a linear system of optical elements that may include lenses and mirrors, and arrive at their respective detection planes. At one detection plane, the incident field illuminates a thin transmission mask, whose spatial transmissivity is the pattern to be measured, and is subsequently detected by a bucket detector that provides no transverse spatial resolution. At the other detection plane, the incident field, which has never interacted with the transmission mask, is detected by a pinhole detector centered at some transverse coordinate $\rhovec_{1}$.  The two photocurrents are then correlated and the output value is registered.  This process is repeated as the pinhole detector is scanned along the transverse plane.  The resulting correlation measurements, when viewed as a function of $\rhovec_1$, reveals the power transmissivity of the mask. The image obtained by this procedure has been called a ``ghost image," because the bucket detector that captures the optical field which illuminated the transmission mask has no spatial resolution, and the the pinhole detector measures a field that never interacted with the transmission mask~\cite{GIvariations}.

The first demonstration of ghost imaging utilized biphoton-state light obtained from spontaneous parametric downconversion together with photon-counting bucket and pinhole detectors.  This arrangement yielded a background-free image that was interpreted as a quantum phenomenon, owing to the entanglement of the source photons \cite{Pittman}. However, subsequent experimental \cite{Valencia,Ferri} and theoretical \cite{Gatti:three,Gatti} considerations demonstrated that ghost imaging can be performed with thermalized laser light, utilizing either photon-counting detectors or CCD detector arrays to obtain ghost images, albeit with a background. 

The theory of biphoton ghost imaging requires quantum descriptions for both the optical source and its photodetection statistics, whereas thermal-light ghost imaging admits to a semiclassical description employing classical fields and shot-noise limited detectors. This disparity has sparked interest \cite{Bennink:two,DAngelo,Cai,Scarcelli} in establishing a unifying theory that characterizes the fundamental physics of ghost imaging and delineates the boundary between classical and quantum behavior.  In this paper we develop that unifying theory within the framework of Gaussian-state (classical and nonclassical) sources.  

The foundation of our work is laid in Sec.~\ref{analysis}.   Here we begin by expressing the photocurrent cross-correlation---the ghost image---as a filtered fourth-moment of the field operators illuminating the detectors.  Next, we briefly review the quantum-optics definition of classical states and specialize that discussion to zero-mean Gaussian states.   Then, using the moment-factoring theorem for zero-mean Gaussian states, we obtain our fundamental expression for the Gaussian-state ghost image in terms of the phase-insensitive and phase-sensitive cross-correlations between the two detected fields, plus a background.  The final part of Sec.~\ref{analysis} sets the stage for detailed understanding of ghost imaging by extending the standard theory of coherence propagation to include phase-sensitive field states.  

Section~\ref{NFvsFFGaussianStates} analyzes ghost imaging performed with three classes of Gaussian-state sources.  We first consider a source possessing the maximum phase-insensitive cross-correlation---as constrained by its auto-correlation functions---but no phase-sensitive cross-correlation.  Such a source always produces a classical state.  Thermal light is of this class.  We also consider a source with the maximum {\em classical} phase-sensitive cross-correlation, given the same auto-correlations as in the previous case, but no phase-insensitive cross-correlation.   Finally, we treat the latter source when its phase-sensitive cross-correlation is the maximum permitted by quantum mechanics.  The low-brightness, low-flux limit of this quantum source is the biphoton state.  Thus these source classes span the experiments reported in \cite{Pittman,Valencia, Ferri} within a unified analytical framework while admitting classical phase-sensitive light as a new possibility.  In Sec.~\ref{contrast} we discuss the image-contrast behavior that is obtained with these sources, and in Sec.~\ref{ObjDetSeparation} we generalize the Fig.~\ref{GI:propagation} configuration to allow for a non-zero separation between the transmission mask and the bucket detector.  We conclude, in Sec.~\ref{discussion}, with a discussion of the ghost-imaging physics that has been revealed by our analysis.

\section{Analysis}
\label{analysis}

\begin{figure}
\begin{center}
\includegraphics[width= 3in]{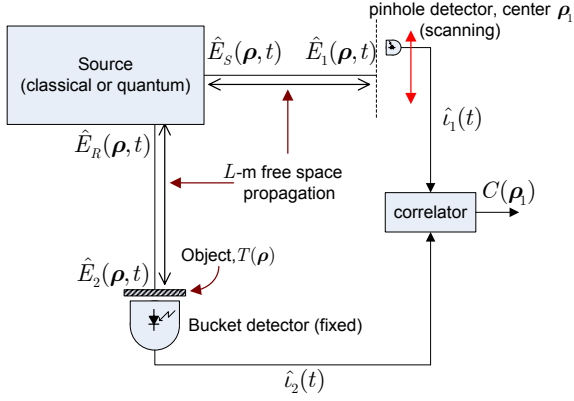}
\end{center}
\caption{(Color online) A simple ghost imaging setup.} \label{GI:propagation}
\end{figure}

Consider the ghost imaging configuration shown, using quantum field and quantum photodetection notation, in Fig.~\ref{GI:propagation}. An optical source generates two optical fields, a signal $\hat{E}_{S}(\rhovec,t)e^{-i \omega_{0}t}$ and a reference $\hat{E}_{R}(\rhovec,t)e^{-i \omega_{0}t}$, that are scalar, positive frequency, paraxial field operators normalized to have units $\sqrt{\text{photons}/{\rm m}^2 {\rm s}}$. Here, $\omega_{0}$ is their common center frequency and $\rhovec$ is the transverse coordinate with respect to each one's optical axis. The commutation relations, within this paraxial approximation, for the baseband field operators are \cite{YuenShapiroPartI}
\begin{align}
[\hat{E}_{m}(\rhovec_{1},t_{1}),\hat{E}_{\ell}(\rhovec_{2},t_{2})]&=0 
 \label{commutators:1}\\
[\hat{E}_{m}(\rhovec_{1},t_{1}),\hat{E}_{\ell}^{\dagger}(\rhovec_{2},t_{2})]&=\delta_{m,\ell}\, \delta(\rhovec_{1} - \rhovec_{2}) \delta(t_{1} - t_{2})\,, \label{commutators:2}
\end{align}
where $\delta_{m, \ell}$ is the Kronecker delta function, $m,\ell = S,R$, and $\delta(\cdot)$ is the unit impulse. Both beams undergo quasimonochromatic paraxial diffraction along their respective optical axes, over an $L$-m-long free-space path, yielding detection-plane field operators \cite{YuenShapiroPartI}
\begin{equation}
\hat{E}_{\ell}(\rhovec,t) \!=\! \int\! {\rm d} \rhovec'\, \hat{E}_{m}\big(\rhovec',t\!-\!L/c \big) h_{L}(\rhovec-\rhovec'), \label{FS:prop}
\end{equation} 
where $(\ell,m) = (1,S) \mbox{ or } (2,R)$, $c$ is the speed of light, $k_{0} = \omega_{0}/c$ is the wave number associated with the center frequency, and $h_{L}(\rhovec)$ is the Huygens-Fresnel Green's function, 
\begin{equation}
h_{L}(\rhovec) \equiv \frac{k_{0} e^{i k_{0} (L+ |\rhovec|^2/2L)}}{i 2 \pi L}\,. \label{HFGreensFunction}
\end{equation}
At the detection planes, $\hat{E}_{1}(\rhovec,t)$ illuminates a quantum-limited pinhole photodetector of area $A_1$ whose photosensitive region $\rhovec \in \mathcal{A}_{1}$ is centered at the transverse coordinate $\rhovec_{1}$, while $\hat{E}_{2}(\rhovec,t)$, illuminates an amplitude-transmission mask $T(\rhovec)$ located immediately in front of a quantum-limited bucket photodetector with photosensitive region $\rhovec \in \mathcal{A}_{2}$. 

\begin{figure}
\begin{center}
\includegraphics[width= 3in]{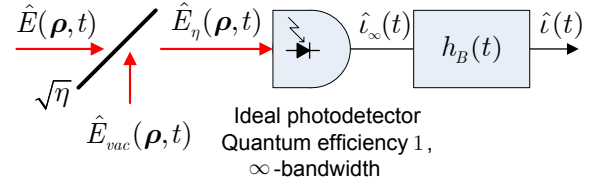}
\end{center}
\caption{(Color online) Photodetection model.} \label{QuantumPhotodetector}
\end{figure}

The photodetectors are assumed to have identical sub-unity quantum efficiencies and finite electrical bandwidths, but no dark current or thermal noise (from subsequent electronics) contributes to the output current. Figure~\ref{QuantumPhotodetector} shows the model utilized for the photodetectors, in which a beam splitter with field-transmissivity $\sqrt{\eta}$ precedes an ideal photodetector to model the real detector's sub-unity quantum efficiency, and a low-pass filter with a real impulse response $h_{B}(t)$ follows the ideal photodetector to model the real detector's finite electrical bandwidth. It follows that the classical output currents from the two detectors correspond to the following quantum measurements \cite{YuenShapiroPartIII,Shapiro:Gaussian,Loudon}: 
\begin{equation}
\hat{\imath}_{m}(t) =  q \! \int\! {\mathrm d}u \int_{\mathcal{A}_{m}} \! {\rm d}\rhovec \,\hat{E}_{\eta,m}^{\dagger}(\rhovec,u) \hat{E}_{\eta,m}(\rhovec,u) h_{B}(t-u), \label{photocurrent}
  \end{equation}
 for $m = 1,2$, 
where $q$ is the electron charge,  
  \begin{equation}
  \hat{E}_{\eta,m}(\rhovec,t) = \sqrt{\eta}\, \hat{E}_{m}(\rhovec, t) + \sqrt{1- \eta}\, \hat{E}_{vac,m}(\rhovec, t)\,, \label{subunityqe}
  \end{equation}
and $\hat{E}_{vac,m}(\rhovec,t)$ is a vacuum-state field operator.  

The ghost image at transverse location $\rhovec_1$ is formed by time-averaging the product of the detector photocurrents to obtain an estimate of the ensemble-average equal-time photocurrent cross-correlation function, which is given by
\begin{eqnarray}
\lefteqn{C(\rhovec_{1}) =  \langle \hat{\imath}_{1}(t) \hat{\imath}_{2}(t) \rangle 
 = q^2 \eta^{2} A_{1} } \nonumber \\[.12in]
 &\times& \int_{\mathcal{A}_{2}}\!{\rm d}\rhovec\int\!{\rm d}u_1 \int\!{\rm d}u_2\, 
    h_{B}(t-u_{1}) h_{B}(t-u_{2}) |T(\rhovec)|^2\nonumber \\[.12in]
 &\times& \langle \hat{E}_{1}^{\dagger}(\rhovec_{1},u_{1}) \hat{E}_{2}^{\dagger}(\rhovec,u_{2}) \hat{E}_{1}(\rhovec_{1},u_{1}) \hat{E}_{2}(\rhovec, u_{2})  \rangle   \,, \label{IntCorr}
\end{eqnarray}
where we have approximated the integral over the pinhole detector's photosensitive region as the value of the integrand at $\rhovec_{1}$ times the photosensitive area, $A_{1}$.

So far we have opted for the quantum description of our ghost imaging configuration, because it applies equally well to both classical-state and nonclassical-state sources.  For the former case, however, we could have arrived at an equivalent answer by use of semiclassical theory. In particular, for a classical-state source we could replace the field operators with scalar classical electromagnetic fields, then employ scalar diffraction theory plus the shot-noise theory for photodetection to arrive at the photocurrent correlation expression in Eq.~ \eqref{IntCorr}, but with the field-operator fourth moment replaced by a classical-field fourth moment. Because a principal goal of this paper is to identify the classical/quantum boundary in ghost imaging, it is incumbent upon us to first review the quantum mechanical states that can generate photocurrent correlations which cannot be obtained from classical electromagnetism and shot-noise theory.  

\subsection{Semiclassical versus quantum photodetection}
\label{semiclVSquantum}

Consider the ideal photodetector from Fig.~\ref{QuantumPhotodetector}, i.e, one with unity quantum efficiency, zero dark current and infinite electrical bandwidth, for which individual photon detection events are registered instantaneously as current impulses carrying charge $q$.  In semiclassical theory, the scalar optical field impinging on the photosensitive surface of the photodetector is a positive-frequency classical electromagnetic wave, denoted by $E (\rhovec,t)e^{-i \omega_{0}t}$. In accordance with the quantum description above, we assume that this field is paraxial, normalized to  have units  $\sqrt{\text{photons}/{\rm m}^2 {\rm s}}$, and has center frequency $\omega_{0}$. Conditioned on knowledge of the field impinging on the photodetector, we have that $i(t)/q$, is an inhomogeneous Poisson impulse train with rate function \cite{Gagliardi,Gallager}
\begin{equation}
\mu(t) = \int_{\mathcal{A}} {\rm d}\rhovec \: \lvert E(\rhovec, t) \rvert^{2} \,,
\end{equation}
where $\mathcal{A}$ is the detector's photosensitive region. Thus, regardless of whether the illuminating field is deterministic or random, the photocurrent is subject to the noise that is inherent in this Poisson process, which yields the well known shot-noise floor of semiclassical photodetection theory \cite{Shapiro:Gaussian}.  Randomness in the illumination is then accounted for by taking $E(\rhovec, t)$ to be a stochastic process, as is done in classical statistical optics \cite{Goodman:Stat}.

In the quantum theory of photodetection, the classical photocurrent produced by the same ideal photodetector 
is a stochastic process whose statistics coincide with those of the photon-flux operator measurement scaled by the electron charge \cite{YuenShapiroPartIII},
\begin{equation}
\hat{\imath} (t) = q \int_{\mathcal{A}} {\rm d}\rhovec \: \hat{E}^{\dagger} (\rhovec,t) \hat{E}(\rhovec,t) \,.
\end{equation}
The photocurrent statistics are then governed by the state of the  field operator $\hat{E}(\rhovec, t)$, so the shot-noise limit of semiclassical theory can be surpassed by some states, such as amplitude-squeezed states, or the eigenkets of continuous-time photodetection \cite{YuenShapiroPartIII,Shapiro:Gaussian,Shapiro:CTEigenkets}.

In our quantum treatment of ghost imaging, the states of the optical field operator $\hat{E}(\rhovec, t)$ that we shall deem {\em classical} are those for which the measurement statistics predicted by quantum photodetection theory match those predicted by the semiclassical theory. It has long been known \cite{YuenShapiroPartIII,Shapiro:Gaussian} that when 
$\hat{E}(\rhovec,t)$ is in the coherent state $|E(\rhovec,t)\rangle$, indexed by its eigenfunction 
$E(\rhovec,t)$ and satisfying
\begin{equation}
\hat{E}(\rhovec, t) \lvert E(\rhovec,t)\rangle  = E(\rhovec,t) \lvert E(\rhovec,t)\rangle\,,
\end{equation}
the statistics of the $\hat{\imath}(t)$ measurement are identical to those from the semiclassical theory with the impinging classical field taken to be $E(\rhovec,t)$.   More generally, the two photodetection theories yield identical statistics for any quantum state that is a classical statistical mixture of coherent states---viz., for all states that have proper $P$-representations \cite{Mandel}---when the classical field used in the semiclassical theory is comprised of the same statistical mixture of the coherent-state eigenfunctions \cite{Shapiro:Gaussian,ShapiroSun,ShapiroYuenMataPartII}. Moreover, mixtures of coherent states are the only quantum states for which all quantum photodetection statistics coincide with the corresponding results found from the semiclassical theory.

The quantum and semiclassical theories of photodetection accord very different physical interpretations to their fundamental noise sources---quantum noise of the illuminating field versus shot noise arising from the discreteness of the electron charge---but for quantum states with proper $P$-representations their predictions are quantitatively indistinguishable. So, because the correlation measurement at the heart of ghost imaging is a derived statistic from two photodetection measurements, the quantum theory of ghost imaging using states that have proper $P$-representations is equivalent to the classical theory of ghost imaging using (classical) random optical fields plus shot-noise detection theory. Therefore, any truly quantum features of ghost imaging must be exclusive to optical field states that do \em not\/\rm\ possess proper $P$-representations.
 
\subsection{Jointly Gaussian states}
\label{GaussianStates}

Gaussian states offer both a practically relevant and a theoretically convenient framework for studying ghost imaging. Their practical relevance stems from thermal states and the biphoton state being special instances of Gaussian states. Their theoretical convenience arises from their being completely determined by their first and second moments, and from their closure under linear transformations.  
Moreover, as noted in Sec.~I, Gaussian-state sources span the experiments reported in 
\cite{Pittman,Valencia,Ferri} and admit to the additional case of classical phase-sensitive light.  Hence they provide an excellent unifying framework within which to probe the distinction between quantum and classical behavior in ghost imaging.

Because the experiments in \cite{Pittman,Valencia, Ferri} employed zero-mean states, we shall assume that $\hat{E}_{S}(\rhovec,t)$ and $\hat{E}_{R}(\rhovec,t)$ are in a zero-mean, jointly Gaussian state, i.e., the characteristic functional of their joint state has a Gaussian form \cite{Shapiro:Gaussian}  specified by the (normally-ordered) phase-insensitive auto- and cross-correlations $\langle \hat{E}_{m}^{\dagger}(\rhovec_{1}, t_{1}) \hat{E}_{\ell}(\rhovec_{2}, t_{2})\rangle$,
and the phase-sensitive auto- and cross-correlations $\langle \hat{E}_{m}(\rhovec_{1}, t_{1}) \hat{E}_{\ell}(\rhovec_{2}, t_{2})\rangle$, 
where $m,\ell = S,R$.  Because the experiments in \cite{Pittman,Valencia, Ferri} employed states whose phase-sensitive auto-correlations were zero, we shall assume that $\langle \hat{E}_{m}(\rhovec_{1}, t_{1}) \hat{E}_{m}(\rhovec_{2}, t_{2})\rangle = 0$ for $m = S, R$.
Finally, to simplify our analytical treatment, while preserving the essential physics of ghost imaging, we shall assume that the signal and reference fields are cross-spectrally pure, complex-stationary and have identical auto-correlations, i.e.,
\begin{align}
\!\langle \hat{E}_{m}^{\dagger}(\rhovec_1,t_1) \hat{E}_{m}(\rhovec_2,t_2) \rangle &\!=\! K^{(n)}(\rhovec_1, \rhovec_2) R^{(n)}(t_2\!-\!t_1) \label{SourceCORR:PISAC}\\
\!\langle \hat{E}_{S}^{\dagger}(\rhovec_1,t_1) \hat{E}_{R}(\rhovec_2,t_2) \rangle &\!=\! K^{(n)}_{S,R}(\rhovec_1, \rhovec_2) R^{(n)}_{S,R}(t_2\!-\!t_1) \label{SourceCORR:PISCC}\\
\!\langle \hat{E}_{S} (\rhovec_1,t_1) \hat{E}_{R}(\rhovec_2,t_2) \rangle  &\!=\! K^{(p)}_{S,R}(\rhovec_1, \rhovec_2) R^{(p)}_{S,R} (t_2\!-\!t_1) \label{SourceCORR:PS}
\end{align}
for $m = S,R$, where the superscripts $(n)$ and $(p)$  label normally-ordered (phase-insensitive) and phase-sensitive terms, respectively.  For convenience, and with no loss of generality, we shall assume that 
\begin{equation}
R^{(n)}(0) = R^{(n)}_{S,R}(0) = R^{(p)}_{S,R}(0) = 1.
\end{equation}

With the exception of the behavior of a  background term, the physics of ghost imaging will be shown to arise entirely from the spatial terms in the preceding correlation functions.  These will be taken to have Schell-model forms \cite{Mandel},
\begin{align}
K^{(n)}(\rhovec_1, \rhovec_2) & = A^{*}(\rhovec_1) A(\rhovec_2) G^{(n)}(\rhovec_2-\rhovec_1)\\ 
K^{(n)}_{S,R}(\rhovec_1, \rhovec_2) & = A^{*}(\rhovec_1) A(\rhovec_2) G^{(n)}_{S,R}(\rhovec_2-\rhovec_1)\\ 
K^{(p)}_{S,R}(\rhovec_1, \rhovec_2)  &= A(\rhovec_1) A(\rhovec_2) G_{S,R}^{(p)}(\rhovec_2-\rhovec_1)\,,
\end{align}
with $|A(\rhovec)| \leq 1$, so that this function may be regarded as a (possibly complex-valued) pupil function that truncates a statistically homogeneous random field with phase-insensitive auto-correlations $G^{(n)}(\rhovec_{2}-\rhovec_{1})$, phase-insensitive cross-correlation $G^{(n)}_{S,R} (\rhovec_{2}-\rhovec_{1})$, and phase-sensitive cross-correlation $G^{(p)}_{S,R}(\rhovec_{2}-\rhovec_{1})$.  We shall also assume that $G^{(n)}(\rhovec)$ is a real-valued even function of its argument \cite{coherence_separability}.  Our task, in the rest of this subsection, is to establish the correlation-function bounds that distinguish between classical and quantum behavior for the preceding jointly Gaussian states.  

Let us begin with Gaussian-state signal and reference fields that have only phase-insensitive correlations, i.e., assume that $\langle \hat{E}_{S} (\rhovec_1,t_1) \hat{E}_{R}(\rhovec_2,t_2) \rangle = 0$. Then, the phase-insensitive correlation spectra, given by the three-dimensional Fourier transforms 
\begin{align}
\tilde{g}^{(n)}(\kvec,\Omega) &\equiv \mathcal{F} \left \{ G^{(n)}(\rhovec)R^{(n)}(\tau) \right \} \\
\tilde{g}^{(n)}_{S,R}(\kvec,\Omega) &\equiv \mathcal{F} \left \{ G^{(n)}_{S,R}(\rhovec)R^{(n)}_{S,R}(\tau) \right \}, 
\end{align} 
must satisfy \cite{Shapiro:Gaussian,ShapiroSun} the Cauchy-Schwarz inequality,
\begin{equation}
\lvert \tilde{g}^{(n)}_{S,R}(\kvec,\Omega) \rvert \leq \tilde{g}^{(n)}(\kvec,\Omega)\,, \label{CS:PISq}
\end{equation}
from stochastic process theory \cite{Papoulis}.  Because the correlation spectra in \eqref{CS:PISq} fully determine the zero-mean, phase-insensitive, Gaussian state we are considering, this inequality is both necessary and sufficient to conclude (via the equivalence developed in Sec.~\ref{semiclVSquantum}) that all phase-insensitive Gaussian states have proper $P$-representations, and are therefore classical~\cite{PIS:proofsketch}. The 50/50 beam splitting of a continuous-wave laser beam that has first been transmitted through a rotating ground-glass diffuser---as was done in the experiments of \cite{Valencia,Ferri}---yields signal and reference fields that are in a zero-mean, phase-insensitive, jointly Gaussian state in which \eqref{CS:PISq} is satisfied with equality.

Now let us examine the more interesting case in which the zero-mean Gaussian-state signal and reference fields have a non-zero phase-sensitive cross-correlation, but no phase-insensitive cross-correlation.  Here we will find that their joint state need {\em not} have a proper $P$-representation, viz., the state may be nonclassical. We have that the phase-sensitive cross-correlation spectrum of the signal and reference fields,
\begin{equation}
\tilde{g}^{(p)}_{S,R}(\kvec,\Omega) \equiv \mathcal{F} \left \{ G^{(p)}_{S,R}(\rhovec)R^{(p)}_{S,R} (\tau) \right \}\,,
\end{equation}
satisfies \cite{Shapiro:Gaussian,ShapiroSun}
\begin{equation}
\lvert \tilde{g}^{(p)}_{S,R}(\kvec,\Omega) \rvert \leq \sqrt{\left [1 + \tilde{g}^{(n)}(\kvec,\Omega)\right] \tilde{g}^{(n)}(\kvec,\Omega)}\,, \label{CS:PSq}
\end{equation}
whereas the Cauchy-Schwarz inequality for the phase-sensitive cross-correlation spectrum of a pair of classical stochastic processes imposes the more restrictive condition \cite{Shapiro:Gaussian,ShapiroSun}
\begin{equation}
\lvert \tilde{g}^{(p)}_{S,R}(\kvec,\Omega) \rvert \leq \tilde{g}^{(n)}(\kvec,\Omega)\,. \label{CS:PSc}
\end{equation}
Zero-mean Gaussian states whose phase-sensitive cross-correlation spectra satisfy \eqref{CS:PSc} have characteristic functionals consistent with that of a pair of classical stochastic processes.  Hence these states have proper $P$-representations and are therefore classical. On the other hand, zero-mean Gaussian states whose phase-sensitive cross-correlation spectra violate \eqref{CS:PSc} have characteristic functionals that are inadmissible in stochastic process theory and are therefore nonclassical. In short, equality in \eqref{CS:PSc} constitutes a well-defined boundary between classical and nonclassical zero-mean Gaussian states.

The difference between inequalities  \eqref{CS:PSq} and \eqref{CS:PSc} has a simple physical origin.  Both derive from the fact that linear combinations of signal and reference fields have non-negative measurement variances.  In the quantum case, however, the variance calculation leading to \eqref{CS:PSq} must invoke the field-operator commutators, whereas the derivation of \eqref{CS:PSc} has no such need.  (Note that commutator issues do \em not\/\rm\ arise in deriving \eqref{CS:PISq}, which is why this inequality is the same for the quantum and classical cases.)  
The upper bounds in \eqref{CS:PSq} and \eqref{CS:PSc} are similar for $\tilde{g}^{(n)}(\kvec,\Omega) \gg 1$.  Thus it might seem that there is little distinction between classical and quantum Gaussian states in this limit.   While this will be seen below to be so for ghost imaging (when background is neglected), 50/50 linear combinations of the signal and reference fields will be highly squeezed---thus highly nonclassical---when  $\tilde{g}^{(n)}(\kvec,\Omega) \gg 1$.  At the other extreme, for $\tilde{g}^{(n)}(\kvec,\Omega) \ll 1$, the quantum upper bound is approximately $\sqrt{\tilde{g}^{(n)}(\kvec,\Omega)}$, which is significantly greater than the classical upper bound $\tilde{g}^{(n)}(\kvec,\Omega)$.  The phase-insensitive correlation spectrum $\tilde{g}^{(n)}(\kvec,\Omega)$ specifies the brightness of the signal and idler fields in units of photons.  Thus $\tilde{g}^{(n)}(\kvec,\Omega) \ll 1$ is a low-brightness condition.   
In this regime we will see that there are appreciable differences between the ghost image formed with classical phase-sensitive light and quantum phase-sensitive light.  

Spontaneous parametric downconversion (SPDC), which was used in the original ghost imaging experiment \cite{Pittman}, produces signal and reference fields that are in a zero-mean jointly Gaussian state with no phase-insensitive cross-correlation and no phase-sensitive auto-correlation, but with a phase-sensitive cross-correlation that saturates the upper bound in \eqref{CS:PSq}. Furthermore, for continuous-wave SPDC operating at frequency degeneracy, this state is a two-field minimum-uncertainty-product pure state, generated by the Bogoliubov transformation  \cite{Brambilla:SPDC,WongKimShapiro}
\begin{align}
\nonumber \hat{E}_{S}(\kvec,\Omega) & = M(\kvec, \Omega)  \hat{E}_{S_v}(\kvec,\Omega) \\ &\hspace*{12pt} + V(\kvec, \Omega) \hat{E}_{R_v}^{\dagger}(-\kvec,-\Omega)\,, \\
\nonumber \hat{E}_{R}(-\kvec,-\Omega) & = M(\kvec, \Omega) \hat{E}_{R_v}(-\kvec,-\Omega) \\ &\hspace*{12pt} + V(\kvec, \Omega) \hat{E}_{S_v}^{\dagger}(\kvec,\Omega)\,,
\end{align}
of the vacuum-state input fields, $\hat{E}_{S_v}$ and $\hat{E}_{R_v}$, where the transfer functions satisfy $\lvert M(\kvec, \Omega) \rvert^{2} - \lvert V(\kvec, \Omega) \rvert^{2} =1 $ to preserve the free-field commutator relations given in \eqref{commutators:1} and \eqref{commutators:2}.

In the low-brightness, low-flux regime, wherein $g^{(n)}(\kvec, \Omega) \ll 1$, $\lvert g^{(p)}(\kvec, \Omega) \rvert \approx \sqrt{\tilde{g}^{(n)}(\kvec,\Omega)}$ and at most one signal-reference photon pair is present in the electrical time constant of the ghost imager's photodetectors, the first and second-order moments of this Gaussian state match those obtained from the unnormalizable pure state comprised of a superposition of a dominant multimode vacuum and a weak biphoton component \cite{WongKimShapiro}, i.e.,
\begin{eqnarray}
|\psi\rangle &=& | \mathbf{0} \rangle_S |\mathbf{0}\rangle_R + \int\! {\rm d}\kvec\int\!{\rm d}\Omega \, 
\nonumber \\[.12in] &\times& \sqrt{\tilde{g}^{(n)}(\kvec,\Omega)}\, e^{i \phi(\kvec,\Omega)} |\kvec , \Omega \rangle_S \,  |-\kvec , -\Omega \rangle_R \,, \label{biphoton}
\end{eqnarray}
where $\phi(\kvec, \Omega)  \equiv \angle g^{(p)}(\kvec, \Omega)$ and $| \mathbf{0} \rangle_S |\mathbf{0} \rangle_R$ is the multimode vacuum state of the signal and reference.  In the biphoton term,  $|\kvec,\Omega\rangle_S$ denotes the single-photon signal-field state with transverse wave vector $\kvec$ and frequency detuning $\Omega$ from degeneracy; a similar interpretation applies to the reference-field state $|-\kvec,-\Omega\rangle_R$.  So, because the pure state given in  Eq.~\eqref{biphoton} is the low-brightness, low-flux equivalent of the zero-mean jointly Gaussian state with maximum phase-sensitive cross-correlation and no phase-insensitive cross-correlation, it is clear that Gaussian-state analysis encompasses the previous biphoton treatments of ghost imaging using SPDC.  

As a final point about jointly Gaussian states,  let us note how one may obtain classical phase-sensitive cross-correlations between the signal and reference fields.  Such fields can be generated by imposing complex-conjugate zero-mean Gaussian-noise modulations, in space and time, on the fields obtained by 50/50 beam splitting of a continuous-wave laser beam. This saturates the upper bound in \eqref{CS:PSc}, because the resulting joint state is a Gaussian statistical mixture of the coherent states $|E(\rhovec,t) \rangle_{S} | E^{*}(\rhovec,t) \rangle_{R}$.  Existing modulator technology will limit the bandwidth achievable with such an arrangement to tens of GHz.  Substantially broader bandwidths might be realized by exploiting the classical (high-photon-flux) limit of nonlinear processes that generate phase-conjugate beams \cite{Boyd}.

\subsection{Coherence propagation}
\label{CoherenceProp}

The previous subsection laid out the statistical source models that we shall employ in our ghost imaging analysis; it was grounded in the second moments of the source-plane field operators $\hat{E}_{S}$ and $\hat{E}_{R}$ that completely characterize their zero-mean, jointly Gaussian state.  However, our expression for the photocurrent correlation $C(\rhovec_1)$ in the 
Fig.~\ref{GI:propagation} ghost-imaging configuration is given by Eq.~\eqref{IntCorr}, which requires a fourth moment of the detection-plane field operators $\hat{E}_{1}$ and $\hat{E}_{2}$.  These detection-plane operators result from $L$\,m free-space propagation of the source-plane operators, as given by Eq.~\eqref{FS:prop}.   Jointly Gaussian states remain jointly Gaussian under linear transformations, such as Eq.~\eqref{FS:prop}, and zero-mean states remain zero-mean as well.  Thus, free-space diffraction over the $L$-m-long propagation paths transform the zero-mean, jointly Gaussian state of the source, with correlation functions given in Eqs.~\eqref{SourceCORR:PISAC}--\eqref{SourceCORR:PS}, into a zero-mean, jointly Gaussian state at the detection planes whose correlation functions are cross-spectrally pure and given by 
\begin{align}
\!\langle \hat{E}_{m}^{\dagger}(\rhovec_1,t_1) \hat{E}_{\ell}(\rhovec_2,t_2) \rangle \!&=\! K^{(n)}_{m,\ell}(\rhovec_1, \rhovec_2) R^{(n)}_{m,\ell}(t_2-t_1) \label{DetCORR:PIS}\\ 
\!\langle \hat{E}_{1} (\rhovec_1,t_1) \hat{E}_{2}(\rhovec_2,t_2) \rangle  \!&=\! K^{(p)}_{1,2}(\rhovec_1, \rhovec_2) R^{(p)}_{1,2} (t_2-t_1). \label{DetCORR:PS}
\end{align}
In these expressions, 
\begin{align}
\nonumber K^{(n)}_{m,\ell}(\rhovec_1, \rhovec_2) = \int\! & {\rm d}\rhovec_1' \int\!{\rm d}\rhovec_2' \: K^{(n)}_{m',\ell'}(\rhovec_1', \rhovec_2') \\ & \times h_{L}^{*}(\rhovec_{1}-\rhovec_{1}') h_{L}(\rhovec_{2}-\rhovec_{2}')\,,  \label{FSProp:PIS}\\
\nonumber K^{(p)}_{1,2}(\rhovec_1, \rhovec_2) = \int\! & {\rm d}\rhovec_1'\int\! {\rm d}\rhovec_2' \: K^{(p)}_{S,R}(\rhovec_1', \rhovec_2') \\ & \times h_{L}(\rhovec_{1}-\rhovec_{1}') h_{L}(\rhovec_{2}-\rhovec_{2}') \,,\label{FSProp:PS}
\end{align}
for $(m,m') = (1,S)$ or $(2,R)$, and likewise for $(\ell, \ell')$. Also, the temporal correlation behavior is unaffected by propagation, because the quasimonochromatic quantum Huygens-Fresnel principle, Eq.~\eqref{FS:prop}, only involves delay in time.   It follows that the fundamental difference between the propagation of phase-insensitive and phase-sensitive correlation functions is the lack of conjugation in the propagation kernel of the latter, something which is responsible for the propagation characteristics reviewed in Sec.~\ref{NFvsFF:Propagation} \cite{ErkmenShapiro:PSCoherenceThy}.

Previous work on biphoton imaging has shown that the biphoton state propagates through free space in the same manner shown above for the phase-sensitive cross-correlation function  \cite{Saleh:Duality}.  This is  \em not\/\rm\ coincidental.   We know from Eq.~\eqref{biphoton} that the biphoton wave function is the phase-sensitive cross-correlation between signal and reference fields with maximum phase-sensitive cross-correlation in the low-brightness, low-flux limit.  For more general Gaussian states---which can have arbitrary brightness and photon flux and can be classical or nonclassical---it is necessary to consider the propagation of the phase-sensitive cross-correlation function.

Having related second moments of the detection-plane fields to their source-plane counterparts, we still need to find a fourth moment of those detection-plane fields in order to evaluate Eq.~\eqref{IntCorr}.  For zero-mean jointly Gaussian states this step is easy.  From the Gaussian moment-factoring theorem \cite{Mandel} we find that the fourth-order moment in Eq.~\eqref{IntCorr} is given by
\begin{eqnarray}
\lefteqn{\hspace*{-.2in}\langle \hat{E}_{1}^{\dagger}(\rhovec_{1},u_{1}) \hat{E}_{2}^{\dagger}(\rhovec,u_{2}) \hat{E}_{1}(\rhovec_{1},u_{1}) \hat{E}_{2}(\rhovec,u_{2})  \rangle = } \nonumber \\[.12in]
&&\hspace*{-.3in} \langle \hat{E}_{1}^{\dagger}(\rhovec_{1},u_{1}) \hat{E}_{1}(\rhovec_{1},u_{1}) \rangle \langle \hat{E}_{2}^{\dagger}(\rhovec,u_{2}) \hat{E}_{2}(\rhovec,u_{2})  \rangle + \nonumber \\[.12in] &&
\hspace*{-.3in}  \lvert \langle \hat{E}_{1}^{\dagger}(\rhovec_{1},u_{1})   \hat{E}_{2}(\rhovec,u_{2}) \rangle \rvert^{2} + \lvert \langle \hat{E}_{1}(\rhovec_{1},u_{1})   \hat{E}_{2}(\rhovec,u_{2}) \rangle \rvert^{2}. \label{GMFT}
\end{eqnarray}
Substituting Eq.~\eqref{GMFT} into Eq.~\eqref{IntCorr}, along with Eqs.~\eqref{DetCORR:PIS} and \eqref{DetCORR:PS}, simplifies the photocurrent cross-correlation expression to 
\begin{eqnarray}
C(\rhovec_{1}) &=& C_{0}(\rhovec_{1}) + C_{n} \int_{\mathcal{A}_{2}}\! {\rm d}\rhovec\,  |K^{(n)}_{1,2}(\rhovec_1, \rhovec)|^{2}  |T(\rhovec)|^{2} \nonumber \\[.05in] &+& C_{p}  \int_{\mathcal{A}_{2}}\! {\rm d}\rhovec\,  |K^{(p)}_{1,2}(\rhovec_1, \rhovec)|^{2}  |T(\rhovec)|^{2}\,, \label{GI:corr}
\end{eqnarray}
where
\begin{eqnarray}
\lefteqn{C_{0}(\rhovec_{1}) = q^2 \eta^2 A_{1} R^{(n)}_{1,1}(0) R^{(n)}_{2,2}(0) \left (\int\! h_{B}(t) {\rm d}t \right)^{2} }\nonumber \\[,12in] &\times& K^{(n)}_{1,1}(\rhovec_1, \rhovec_1) \int_{\mathcal{A}_{2}} \! {\rm d}\rhovec \: K^{(n)}_{2,2}(\rhovec, \rhovec) \lvert T(\rhovec) \rvert^{2} \,, \label{Co}
\end{eqnarray}
is a non-negative non-image-bearing background, and 
\begin{align}
C_{n} & = q^2 \eta^2 A_{1} \left [ \lvert R^{(n)}_{1,2}(t)   \rvert^{2}  \star h_{B}(t) \star h_B(-t) \right ]_{t=0}\,, \label{Cn}\\
C_{p} & = q^2 \eta^2 A_{1} \left [ \lvert R^{(p)}_{1,2}(t)   \rvert^{2}  \star h_{B}(t) \star h_B(-t) \right ]_{t=0} \label{Cp} \,,
\end{align}
are constants that depend on the temporal cross-correlations between $\hat{E}_{1}$ and $\hat{E}_{2}$. Here $\star$ denotes convolution. 

The image-bearing term in $C(\rhovec_{1})$ is seen, from Eq.~\eqref{GI:corr}, to be the object's intensity transmission profile, $|T(\rhovec)|^2$, filtered through a linear, space-varying filter whose point-spread function is given by a weighted sum of the squared magnitudes of the phase-insensitive and phase-sensitive cross-correlation functions at the detection planes.
In thermal-state ghost imaging, the phase-sensitive term vanishes, so that the point-spread function depends only on the phase-insensitive cross-correlation. In biphoton-state ghost imaging, the phase-insensitive cross-correlation is zero, thus yielding an image filter that depends only on the phase-sensitive cross-correlation. For general Gaussian-state signal and reference fields, however, both cross-correlations contribute to the image filter.

Because the image-bearing part of Eq.~\eqref{GI:corr}  only depends on the cross-correlations between the detected fields, whereas the non-image-bearing background depends only on the phase-insensitive auto-correlations, it is germane to note (see Appendix~\ref{appendix}) that {\em any} pair of phase-insensitive and phase-sensitive cross-correlation functions can be associated with a {\em classical} zero-mean jointly Gaussian state, by appropriate choices of its phase-insensitive auto-correlation functions. Thus, if no constraint is placed on the background level in which the image is embedded, i.e., if the auto-correlation functions are not constrained, any image-bearing term attainable from Eq.~\eqref{GI:corr} with a nonclassical Gaussian state source can be replicated {\em identically} by a classical Gaussian-state source. Hence, ghost-image formation is intrinsically classical.  

\subsection{Near-field versus far-field propagation}
\label{NFvsFF:Propagation}

Here  we review the main results for paraxial, quasimonochromatic, phase-insensitive and phase-sensitive coherence propagation through free space \cite{ErkmenShapiro:PSCoherenceThy} that will be combined, in the next section, with Eq.~\eqref{GI:corr} to identify the imaging properties of the Fig.~\ref{GI:propagation} configuration.   Because Eqs.~\eqref{FSProp:PIS} and \eqref{FSProp:PS} show that propagation only affects the correlation functions' spatial components, we shall focus exclusively on them. In order to highlight the difference between the propagation of phase-insensitive and phase-sensitive coherence, we shall consider (real and even) Gaussian-Schell model spatial cross-correlation functions for both, i.e., we assume \cite{autoVScross}
\begin{equation}
K^{(x)}_{S,R}(\rhovec_1, \rhovec_2) = \frac{2P}{\pi a_{0}^2} e^{-(|\rhovec_{1}|^2+ |\rhovec_{2}|^2)/a_{0}^2 - |\rhovec_{2} - \rhovec_{1}|^2/2 \rho_{0}^2 } \label{NearVsFar}\,,
\end{equation}
for $x =  n, p$, where $P$ is the photon flux of the signal (and idler), $a_{0}$ is the $e^{-2}$ attenuation radius of the transverse intensity profile, and $\rho_{0}$ is the transverse coherence radius, which is assumed to satisfy the low-coherence condition $\rho_{0} \ll a_{0}$.

We compare phase-insensitive and phase-sensitive correlation propagation in two limiting regimes: the near field, which corresponds to the region in which diffraction effects are negligible, and the far field, in which diffraction spread is dominant. For phase-insensitive coherence propagation, it is well known that a single Fresnel number, $D_{0} = k_{0} \rho_{0} a_{0} / 2 L$, distinguishes between these regimes, with $D_{0}\gg 1$ corresponding to the near field and $D_{0} \ll 1$ being the far field \cite{Goodman:Stat,Mandel}. Note that this Fresnel number differs from that for the diffraction of a coherent laser beam with intensity radius $a_0$, which is $D_{coh} = k_{0} a^{2}_{0} / 2 L$.  This difference reflects the coupling between coherence radius and intensity radius that occurs in free-space diffraction of partially-coherent light. In particular, far-field propagation of the phase-insensitive correlation function from Eq.~\eqref{NearVsFar} results in an intensity radius satisfying $a_{L} = a_{0}/D_{0} = 2L/k_0\rho_0$ and a coherence radius given by $\rho_{L} = \rho_{0}/D_{0} = 2L/k_0a_0$, i.e., the far-field intensity radius is inversely proportional to its source-plane coherence length and the far-field coherence length is inversely proportional to the source-plane intensity radius.

The phase-sensitive correlation function from Eq.~\eqref{NearVsFar} propagates in a distinctly different manner from its phase-insensitive counterpart.  In this case we find that coherence-radius diffraction and intensity-radius diffraction are decoupled \cite{ErkmenShapiro:PSCoherenceThy}.  Two Fresnel numbers are then required to distinguish the near field from the far field:  the Fresnel number for diffraction of the coherence length, $D_{N} = k_{0} \rho_{0}^{2}/2 L$; and the Fresnel number for diffraction of the intensity radius, $D_{F} = k_{0} a_{0}^2 /2 L$. The near-field regime for phase-sensitive correlation propagation occurs when {\em both} Fresnel numbers are much greater than one, and the far-field regime is when both are much less than one.  Because we have imposed the low-coherence condition, $\rho_{0} \ll a_{0}$,  we can say that the near-field regime for phase-sensitive coherence propagation is $D_{N} \gg 1$ and its far-field regime is $D_{F} \ll 1$.   Each of these conditions is  more stringent than the corresponding condition for phase-insensitive light. Nevertheless, the far-field propagation of the Gaussian-Schell model phase-sensitive correlation function from \eqref{NearVsFar} still yields $\rho_{0}/D_{0}$ for the far-field intensity radius and $a_{0}/D_{0}$ for the far-field coherence radius.  However, whereas the far-field phase-insensitive correlation is highest for two points with equal transverse-plane coordinates, the far-field phase-sensitive correlation is highest for two points that are symmetrically disposed about the origin on the transverse plane \cite{ErkmenShapiro:PSCoherenceThy,Saleh:Duality}.

\begin{figure}[ht]
\centering
\includegraphics[width=3in]{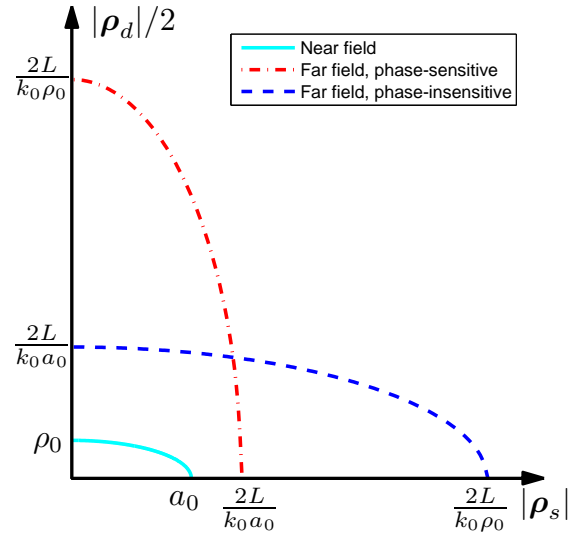}
\caption{(Color online) Isocontours corresponding to the $e^{-2}$-attenuation levels for the phase-sensitive and phase-insensitive correlation functions in the near-field and the far-field regimes.}
\label{GS:LC}
\end{figure}

Figure~\ref{GS:LC} highlights the difference between propagation of the phase-insensitive and phase-sensitive correlation functions from Eq.~\eqref{NearVsFar}.  In this figure we have plotted the $e^{-2}$-attenuation isocontours for $|K_{S,R}^{(x)}(\rhovec_1,\rhovec_2)|$, for $x = n,p$, versus the sum and difference coordinates $\rhovec_{s} \equiv (\rhovec_{2} + \rhovec_{1})/2$ and $\rhovec_{d} \equiv \rhovec_{2} - \rhovec_{1}$. All transverse-coordinate pairs that correspond to the interior region of a contour are both coherent and intense. From Eq.~\eqref{NearVsFar}, it is straightforward to verify that all  magnitude isocontours of our Gaussian-Schell model correlations are ellipses. In the near field, because of our low-coherence assumption, the $e^{-2}$-attenuation isocontours---for both the phase-insensitive and the phase-sensitive correlation functions---have their minor axes along the difference coordinate. In the far field, we find that diffraction leads to identical increases along the major and minor axes of the phase-insensitive correlation's $e^{-2} $-attenuation isocontour.  For the corresponding far-field phase-sensitive correlation's isocontour we get inverted behavior, with its minor axis now aligned with the sum coordinate and its major axis along the difference coordinate. Thus, the far-field phase-insensitive correlation function is dominated by a narrow function in the difference coordinate $| \rhovec_d |$, whereas the far-field phase-sensitive correlation function is a narrow function in the sum coordinate $| \rhovec_s |$. 

\section{Near- and Far-Field Ghost Imaging with Gaussian-State Light}
\label{NFvsFFGaussianStates}

We are now fully equipped to compare the ghost-imaging performance achieved in the Fig.~\ref{GI:propagation} configuration with various Gaussian-state sources.  We shall assume that the signal and reference fields $\hat{E}_S$ and $\hat{E}_R$ are in a zero-mean, jointly Gaussian state with identical phase-insensitive auto-correlations given by the following Gaussian-Schell model: 
\begin{eqnarray}
\lefteqn{\hspace*{-.2in}K^{(n)}(\rhovec_1, \rhovec_2) R^{(n)}(t_2-t_1)= } \nonumber \\[.12in]
&& \hspace*{-.2in}\frac{2 P}{\pi a_{0}^2} e^{-(|\rhovec_{1}|^2+ |\rhovec_{2}|^2)/a_{0}^2 - |\rhovec_{2} - \rhovec_{1}|^2/2 \rho_{0}^2 } e^{-(t_2 - t_1)^2/2T_{0}^{2}} \label{GSM:PIS}\,,
\end{eqnarray}
where $\rho_{0} \ll a_{0}$, $T_{0}$ is the coherence time, and $P$ is the photon flux. We will begin our treatment with the thermal-state source, for which the signal and reference have a non-zero phase-insensitive cross-correlation, but no phase-sensitive cross-correlation.  As noted in Sec.~\ref{GaussianStates}, such states are always classical.

\subsection{Ghost imaging with phase-insensitive light}

Consider jointly Gaussian signal and reference fields with auto-correlations given by Eq.~\eqref{GSM:PIS} and no phase-sensitive auto- or cross-correlations.   Inequality~\eqref{CS:PISq} implies that $\lvert \langle \hat{E}_{S}^{\dagger}(\rhovec_{1},t_1) \hat{E}_{R}(\rhovec_{2},t_2)\rangle \rvert $ is maximum when it equals the auto-correlation function Eq.~\eqref{GSM:PIS}.  We will take this to be so---to maximize the strength of the ghost image---and assume that this phase-insensitive cross-correlation is real-valued and non-negative.  Because near-field detection-plane correlations coincide with source-plane correlations, we can now obtain the near-field ghost image by substituting the right-hand side of Eq.~ \eqref{GSM:PIS} into Eq.~\eqref{GI:corr}.   Doing so gives us the following result, 
\begin{eqnarray}
C(\rhovec_{1}) &=& C_0(\rhovec_{1}) + C_{n} (2 P/\pi a_{0}^2 )^{2} \, e^{-2 |\rhovec_{1}|^2/a^{2} _{0}} 
\nonumber \\[.12in]   &\times& \int_{\mathcal{A}_{2}}\! {\rm d}\rhovec \, e^{-|\rhovec_{1} - \rhovec|^2/\rho_{0}^{2}} e^{-2 |\rhovec|^2/a^{2}_{0}} |T(\rhovec)|^{2}\,.
\label{Thermal:NF}
\end{eqnarray}

Equation~\eqref{Thermal:NF} reveals three significant features of the near-field, thermal-state ghost image. First, the ghost image is space-limited by the reference beam's average intensity profile, so that the object must be placed in the field of view $a_{0}$ \cite{FOV}. Second, the useful transverse scanning range of the pinhole detector is restricted to the field of view $a_{0}$. Finally, and most importantly, the finite cross-correlation coherence length $\rho_{0}$ limits the resolution of the image.  When field-of-view limitations can be neglected, the ghost image in Eq.~\eqref{Thermal:NF} is proportional to the convolution of the object's intensity transmission, $|T(\rhovec)|^2$, with the Gaussian point-spread function $e^{-|\rhovec|^2/\rho_0^2}$.  Thus the spatial resolution, defined here as the radius to the $e^{-2}$-level in the point-spread function, is $\sqrt{2}\rho_0$.

Now let us suppose that the ghost image is formed in the far field, when $D_{0} \ll 1$, with the source correlations as assumed for the near-field regime.  In this case  we must first propagate source correlations---given by the right-hand side of Eq.~\eqref{GSM:PIS}---to the detection planes via Eq.~\eqref{FSProp:PIS}.  It turns out that the detection-plane signal and reference fields still have maximum phase-insensitive cross-correlation, 
\begin{eqnarray}
\lefteqn{\hspace*{-.2in}K_{m,\ell}^{(n)}(\rhovec_{1},\rhovec_{2}) R_{m,\ell}^{(n)}(t_2 - t_1) = } \nonumber \\[.12in]
&&\hspace*{-.2in}\frac{2 P}{\pi a_{L}^2} e^{-(|\rhovec_{1}|^2+ |\rhovec_{2}|^2)/a_{L}^2 - |\rhovec_{2} - \rhovec_{1}|^2/2 \rho_{L}^2 } e^{-(t_2 - t_1)^2/2T_{0}^{2}} \label{GSM:PISFF}\,,
\end{eqnarray}
where $m,\ell \in \{1,2\}$, $a_{L} = 2 L / k_{0} \rho_{0}$ and $\rho_{L} = 2 L / k_{0} a_{0}$, and the ghost image signature becomes,
\begin{eqnarray}
C(\rhovec_{1}) &=& C_0(\rhovec_{1}) + C_{n} (2 P/\pi a_{L}^2 )^{2} \, e^{-2 |\rhovec_{1}|^2/a^{2} _{L}}
\nonumber \\[.12in]  & \times& \int_{\mathcal{A}_{2}}\! {\rm d}\rhovec \, e^{-|\rhovec_{1} - \rhovec|^2/\rho_{L}^{2}} e^{-2 |\rhovec|^2/a^{2}_{L}} |T(\rhovec)|^{2}\,.
\label{Thermal:FF}
\end{eqnarray}
Therefore, the far-field field of view increases to $2L/k_0\rho_0$ while the image resolution degrades to $2\sqrt{2} L / k_{0} a_{0}$, but the three conclusions drawn from the near-field image signature Eq.~\eqref{Thermal:NF} remain valid in the far-field regime. Because the resolution of the image degrades with propagation, so long as field of view is not the limiting factor, it is more desirable to place the object in the source's near field.  

\subsection{Ghost imaging with phase-sensitive light}

Now we shall shift our focus to Gaussian-state signal and reference fields that have a non-zero phase-sensitive cross-correlation, but zero phase-insensitive cross-correlation. Applying the Cauchy-Schwarz bound \eqref{CS:PSc} to the Gaussian-Schell model auto-correlations in Eq.~\eqref{GSM:PIS} we find that the maximum $\lvert \langle \hat{E}_{S}(\rhovec_{1},t_1) \hat{E}_{R}(\rhovec_{2},t_2)\rangle \rvert $ for a {\em classical} Gaussian state is also given by Eq.~\eqref{GSM:PIS}.  Similar to what we did for the phase-insensitive case, we shall take the phase-sensitive cross-correlation to achieve its classical magnitude limit and assume that it is real-valued and non-negative.  Then, because the detection-plane cross-correlation equals the source-plane cross-correlation in the near field, we can immediately get the near-field ghost image by substituting the right-hand side of Eq.~\eqref{GSM:PIS} into Eq.~\eqref{GI:corr}, obtaining    
\begin{eqnarray}
C(\rhovec_{1}) &=& C_0(\rhovec_{1}) + C_{p} (2 P/\pi a_{0}^2 )^{2} \, e^{-2 |\rhovec_{1}|^2/a^{2} _{0}}
\nonumber  \\[.12in]  & \times& \int_{\mathcal{A}_{2}}\! {\rm d}\rhovec \, e^{-|\rhovec_{1} - \rhovec|^2/\rho_{0}^{2}} e^{-2 |\rhovec|^2/a^{2}_{0}} |T(\rhovec)|^{2}\,.
\label{PS:NF}
\end{eqnarray}
Equations~\eqref{Cn} and \eqref{Cp} give $C_n = C_p$, for our Gaussian-Schell model source, making the near-field  ghost image formed with classical phase-sensitive light \em identical\/\rm\ to the near-field ghost image formed with phase-insensitive light. 

When the source-to-object separation is in the far-field regime for phase-sensitive coherence propagation, then the source-plane phase-sensitive cross-correlation that gave the preceding near-field ghost image gives rise to the following detection-plane phase-sensitive cross-correlation \cite{ErkmenShapiro:PSCoherenceThy},
\begin{eqnarray}
\lefteqn{\hspace*{-.2in}K_{1,2}^{(p)}(\rhovec_{1},\rhovec_{2}) R_{1,2}^{(p)}(t_2 - t_1) = } \nonumber \\[.12in]
&& \hspace*{-.2in}\frac{2 P}{\pi a_{L}^2} e^{-(|\rhovec_{1}|^2+ |\rhovec_{2}|^2)/a_{L}^2 - |\rhovec_{2} + \rhovec_{1}|^2/2 \rho_{L}^2 } e^{-(t_2 - t_1)^2/2T_{0}^{2}} \label{GSM:PSFF}\,,
\end{eqnarray}
which leads to 
\begin{equation}
\begin{split} C(\rhovec_{1}) & = C_{0}(\rhovec_{1}) + C_{p} (2 P/\pi a_{L}^{2})^2 \, e^{- 2 |\rhovec_{1}|^2/a^{2}_{L}} \\ &\times \int_{\mathcal{A}_{2}} \!{\rm d}\rhovec \, e^{-|\rhovec_{1} + \rhovec|^2/\rho_{L}^{2}} e^{-2 |\rhovec|^2/a^{2}_{L}} |T(\rhovec)|^{2} \,, \end{split}
\label{PHS_CL:FF}
\end{equation}
for the far-field ghost image formed with classical phase-sensitive light.  Again invoking $C_p = C_n$, for our Gaussian-Schell model source, we see that the far-field ghost image formed with classical phase-sensitive light is an inverted version of the corresponding far-field ghost image formed with phase-insensitive light, i.e., it has field of view $a_L$ and spatial resolution $\sqrt{2}\,\rho_L$, as did the phase-insensitive ghost image,  but the phase-sensitive ghost image is proportional to $|T(-\rhovec)|^2 \star e^{-|\rhovec|^2/\rho^{2}_{L}}$ whereas the phase-insensitive ghost image was proportional to $|T(\rhovec)|^2 \star e^{-|\rhovec|^2/\rho^{2}_{L}}$.

Finally, we turn to the ghost image produced using a nonclassical Gaussian state, i.e., one whose phase-sensitive cross-correlation violates \eqref{CS:PSc}.  In what follows we will restrict our attention to two limiting cases in which the phase-sensitive cross-correlation is coherence separable, so that we may utilize the machinery developed earlier in this paper.  In both cases we will take $\langle \hat{E}_S(\rhovec_1,t_1)\hat{E}_R(\rhovec_2t_2)\rangle$ to be real-valued and non-negative with the maximum magnitude permitted by quantum theory.  The limits of interest for this source will be those of    high brightness, $\tilde{g}^{(n)}(\kvec,\Omega) \gg 1$, and low brightness, $\tilde{g}^{(n)}(\kvec,\Omega) \ll 1$ when the source's auto-correlations are given by Eq.~\eqref{GSM:PIS}.  

At high brightness, the distinction between the cross-correlation functions of the quantum and classical phase-sensitive sources becomes insignificant, so that results given above for the ghost image formed with classical phase-sensitive light are excellent approximations for the quantum case.  At low brightness, however, our assumptions yield a phase-sensitive cross-correlation spectrum satisfying
\begin{eqnarray}
\lefteqn{|\tilde{g}^{(p)}(\kvec,\Omega)|\approx \sqrt{\tilde{g}^{(n)}(\kvec,\Omega)}}\, 
\\[.12in]
&=& 2(2 \pi)^{1/4}\, \sqrt{\frac{P T_{0} \rho_{0}^{2}}{a_{0}^2}} \, e^{-\rho_{0}^{2} |\kvec|^2/4} \, e^{-\Omega^{2}T_0^2/4}\,,\label{GSM:PSlowflux} 
\end{eqnarray}
from which we see that the low-brightness regime corresponds to $P T_{0} \rho_{0}^{2} / a_{0}^2 \ll 1$. The source-plane phase-sensitive cross-correlation in this regime is then found to be
\begin{eqnarray}
\lefteqn{\langle \hat{E}_S(\rhovec_1,t_1)\hat{E}_R(\rhovec_2,t_2)\rangle = \left (2/\pi \right )^{1/4} \sqrt{\frac{a_{0}^{2}}{ P T_{0} \rho_{0}^2}} \,\, \times} \nonumber \\[.12in]
&&\frac{2 P}{\pi a_{0}^{2}} e^{ -(|\rhovec_{1}|^2+ |\rhovec_{2}|^2)/a_{0}^{2} - |\rhovec_{2} - \rhovec_{1}|^2/\rho_{0}^{2} } e^{-(t_2 - t_1)^2/T_{0}^{2}}\,. \label{GSM:PS_NCL}
\end{eqnarray} 
Note that \eqref{GSM:PS_NCL} is still a Gaussian-Schell, cross-spectrally pure correlation function, so that in the source's near field we get
\begin{eqnarray}
\lefteqn{\hspace*{-.35in}C(\rhovec_{1}) = C_0(\rhovec_{1}) + \sqrt{\frac{2}{\pi}}\,\frac{a_0^2}{PT_0\rho_0^2}\,C_p \!\left(\frac{2 P}{\pi a_{0}^{2}}\right)^2 e^{-2|\rhovec_1|^2/a_0^2}} \nonumber\\ &\times&   \int_{\mathcal{A}_{2}}\! {\rm d}\rhovec \, 
e^{-2 |\rhovec_{1} - \rhovec|^2/\rho_{0}^{2}} e^{-2|\rhovec|^2/a_0^2} |T(\rhovec)|^{2}\,. 
\label{PHS:NF_NCL}
\end{eqnarray}
This near-field ghost image has the same field of view, $a_{0}$, as the near-field ghost images formed with classical (phase-insensitive or phase-sensitive) light, but its spatial resolution, $\rho_{0}$, is a factor-of-$\sqrt{2}$ better than the spatial resolutions of those classical near-field imagers.  In addition, the quantum case's image-to-background ratio is much higher than those of the classical imagers, because $a_0^2/PT_0\rho_0^2 \gg 1$ in the low-brightness regime.  

The far-field ghost image for the nonclassical source is obtained by propagating its phase-sensitive cross-correlation from Eq.~\eqref{GSM:PS_NCL} to the detector planes and substituting that result into \eqref{GI:corr}.  The result we obtain is
\begin{eqnarray}
C(\rhovec_{1}) & = &C_{0}(\rhovec_1) + \sqrt{\frac{2}{\pi}}\,\frac{a_0^2}{PT_0\rho_0^2}\,C_p\!\left(\frac{P}{\pi a_{L}^{2}}\right)^2 \, e^{- |\rhovec_{1}|^2/a^{2}_{L}} \nonumber \\[.12in] &\times& \int_{\mathcal{A}_{2}} \!{\rm d}\rhovec \, e^{-|\rhovec_{1} + \rhovec|^2/\rho_{L}^{2}} e^{-|\rhovec|^2/a^{2}_{L}} |T(\rhovec)|^{2}.
\label{PHS_NCL:FF}
\end{eqnarray}
Thus, the far-field resolution achieved with the quantum source equals those realized using the  classical sources considered earlier, but the field of view has been increased by a factor of $\sqrt{2}$.   It is worth pointing out that the quantum-enhancement factors---of spatial resolution in the near field and field of view in the far field---derive from the broadening of the weak spectrum, $\tilde{g}^{(n)}(\kvec,\Omega)$, when its square root is taken. That these enhancement factors both equal $\sqrt{2}$ depends, therefore, on our choosing to use a Gaussian-Schell correlation model.  Other correlation functions would lead to different enhancement factors.  Finally, as found above for the near-field case,  the quantum source yields dramatically higher image-to-background ratio in far-field ghost imaging than both its phase-insensitive and phase-sensitive classical counterparts.

\section{Image Contrast}
\label{contrast}

Thus far we have concentrated on the image-bearing terms in the photocurrent correlation from Eq.~\eqref{GI:corr}. These image-bearing terms are embedded in a background $C_{0}(\rhovec_{1})$, which, as we have seen in the preceding section, is much stronger for classical-source ghost imaging than it is for low-brightness quantum-source ghost imaging.
It therefore behooves us to pay some attention to the effect of background on ghost-imaging systems.  For the sake of brevity, we will limit our discussion to the near-field imagers; the far-field cases can be shown to have similar image-contrast issues.  Also, we shall assume that the transmittance pattern being imaged lies well within the field of view of all these ghost imagers, and restrict ourselves to considering the behavior of $C(\rhovec_1)$ in an observation region ${\cal{R}}$ that encompasses the image-bearing terms while satisfying $|\rhovec_1| \ll a_0$.  In this case
\begin{equation}
{\cal{C}} \equiv  \frac{\max_{\cal{R}}[C(\rhovec_1)] - \min_{\cal{R}}[C(\rhovec_1)]}
{C_0({\bf 0})}
\end{equation}
is a meaningful contrast definition.  Its numerator quantifies the dynamic range of the image-bearing terms in the photocurrent correlation $C(\rhovec_1)$, while its denominator is the featureless background that is present within the observation region.  

For analytical convenience, let us take the baseband impulse response $h_B(t)$ to be a Gaussian with $e^{-2}$-attenuation time duration $T_d$, 
\begin{equation}
h_B(t) = e^{-8t^2/T_d^2}\sqrt{8/\pi T_d^2}. 
\end{equation}
The contrast for the classical (phase-sensitive or phase-insensitive) ghost imagers then satisfies
\begin{equation}
{\cal{C}}^{(c)} = {\cal{C}}_s^{(c)}{\cal{C}}_t^{(c)}, 
\end{equation}
where the spatial ($s$) factor is given by
\begin{equation}
{\cal{C}}_s^{(c)} = \frac{\max_{\rhovec_1}[{\cal{I}}_c(\rhovec_1)] - \min_{\rhovec_1}[{\cal{I}}_c(\rhovec_1)]}
{\int_{{\cal{A}}_2}\!{\rm d}\rhovec\,|T(\rhovec)|^2},
\end{equation}
with
\begin{equation}
{\cal{I}}_c(\rhovec_1) \equiv \int_{{\cal{A}}_2}\!{\rm d}\rhovec\,e^{-|\rhovec_1 - \rhovec|^2/\rho_0^2}
|T(\rhovec)|^2,
\end{equation}
being the point-spread degraded image of $|T(\rhovec)|^2$, 
and the temporal ($t$) factor obeys
\begin{equation}
{\cal{C}}_t^{(c)} = 1/\sqrt{1+(T_d/2T_0)^2}.
\end{equation}
Likewise, for the low-brightness  regime quantum imager we find that its contrast, ${\cal{C}}^{(q)}$, factors into the product of a spatial term
\begin{equation}
{\cal{C}}^{(q)}_s = \sqrt{\frac{2}{\pi}}\frac{a_0^2}{PT_0\rho_0^2}\frac{\max_{\rhovec_1}[{\cal{I}}_q(\rhovec_1)] - \min_{\rhovec_1}[{\cal{I}}_q(\rhovec_1)]}
{\int_{{\cal{A}}_2}\!{\rm d}\rhovec\,|T(\rhovec)|^2},
\end{equation}
with
\begin{equation}
{\cal{I}}_q(\rhovec_1) \equiv \int_{{\cal{A}}_2}\!{\rm d}\rhovec\,e^{-2|\rhovec_1 - \rhovec|^2/\rho_0^2}
|T(\rhovec)|^2,
\end{equation}
being its point-spread degraded image of $|T(\rhovec)|^2$, 
times a  temporal term
\begin{equation}
{\cal{C}}_t^{(q)} = 1/\sqrt{1+T_d^2/2T_0^2}.
\end{equation}

The preceding classical and quantum contrast expressions possess interesting and physically significant behavior.  We shall first explore the classical case.  To get at its contrast behavior, we will assume that $T(\rhovec)$ is a binary amplitude mask, as has often been the case in ghost imaging experiments.  It follows that
\begin{equation}
{\cal{C}}^{(c)}_s \approx 
\rho_0^2/A_T \ll1,
\label{resolutionineq}
\end{equation}
where
\begin{equation}
A_T \equiv \int\!{\rm d}\rhovec\,|T(\rhovec)|^2,
\label{area}
\end{equation}
and the inequality in \eqref{resolutionineq} holds because $A_T/\rho_0^2$ is approximately the number of resolution cells in the ghost image.   Combined with the fact that ${\cal{C}}^{(c)}_t \le 1$, Eq.~\eqref{resolutionineq} shows that classical-source ghost imaging \em always\/\rm\ has low contrast according to our contrast definition.  This is why classical-source ghost imaging has been performed with thermalized laser light and has used {\sc ac}-coupling of the photocurrents to the correlator  \cite{Scarcelli}.  Thermalized laser light is a narrowband source, for which $T_d \ll T_0$ so that ${\cal{C}}^{(c)}_t \approx 1$.  The use of {\sc ac}-coupling implies that the correlator is estimating the {\em cross-covariance} between the photocurrents produced by detectors 1 and 2, rather than their cross-correlation.  This ensemble-average cross-covariance is given by $C(\rhovec_1) - C_0({\rhovec_1})$, so it might seem that covariance estimation alleviates all concerns with the background term.  Such is not the case.  Even though the background term does not appear in the photocurrents' cross-covariance, its shot noise and excess noise dictate that a much longer averaging time will be required to obtain an accurate estimate of this cross-covariance function, i.e., to get a high signal-to-noise ratio ghost image.  Now suppose that  classical-source ghost imaging is attempted using broadband light for which $T_d/T_0 \sim 10^3$, corresponding to a THz-bandwidth source and GHz electrical-bandwidth photodetectors.  In comparison with a narrowband classical-source ghost imager of the same photon flux $P$, the broadband imager must use a $10^6$-times longer time-averaging interval to achieve the same signal-to-shot-noise ratio.  

Turning now to the contrast behavior of the low-brightness quantum-source ghost imager,  our assumption of a binary amplitude mask leads to 
\begin{equation}
{\cal{C}}^{(q)}_s \approx \frac{a_0^2}{PT_0A_T} \gg 1/PT_0
\end{equation}
because of our field-of-view assumption.  Thus in broadband, low-brightness, low-flux quantum ghost imaging we find that 
\begin{equation}
{\cal{C}}^{(q)} \gg 1/PT_d \gg 1,
\end{equation}
where the last inequality invokes the low-flux condition.  This is why biphoton sources yield background-free ghost images  \cite{Pittman,Gatti:three,Gatti}, despite SPDC being a broadband process. 

\section{Relay Optics}
\label{ObjDetSeparation}

\begin{figure}[t]
\begin{center}
\includegraphics[width= 3in]{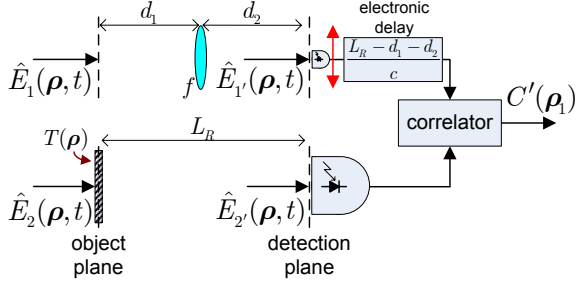}
\end{center}
\caption{(Color online) Ghost imaging setup with relay optics.} \label{GI:detectorplane}
\end{figure}

Our analysis has assumed that the detector plane coincides with the object plane, but a realistic ghost-imaging scenario will likely require a separation between these two planes, as shown in Fig.~\ref{GI:detectorplane}. In this figure, the bucket detector is placed $L_{R}$\,m  away from the object and we assume no control over this path, but we allow ourselves to freely modify the signal-arm path. 
Thus we place a focal-length-$f$ lens $d_{1}$\,m behind the object plane and $d_{2}$\,m in front of the detector plane, such that $1/d_{1} + 1/d_{2} = 1/f$. In addition, because the optical path lengths may be different, we introduce a $(L_{R} - d_{1} - d_{2})/c $ post-detection electronic time delay to maximize the temporal cross-correlation of the two detected fields. The resulting photocurrent cross-correlation is then 
\begin{eqnarray}
C'(\rhovec_{1}) &=& C_{0}(\rhovec_{1}) + C_{n} \int_{\mathcal{A}_2}  d \rhovec_{2}\,   |K^{(n)}_{1',2'}(\rhovec_{1}, \rhovec_{2})|^{2} \nonumber \\[.12in] &+& C_{p} \int_{\mathcal{A}_2}  d \rhovec_{2}\, |K^{(p)}_{1',2'}(\rhovec_{1}, \rhovec_{2})|^{2} \,, \label{GI:corrseparate}
\end{eqnarray}
in terms of the phase-insensitive and phase-sensitive cross-correlations, $K^{(m)}_{1',2'}(\rhovec_{1}, \rhovec_{2})$ for $m = n,p$, of the detected fields $\hat{E}_{1'}$ and $\hat{E}_{2'}$.  
The magnitudes of these detection-plane cross-correlations are easily found from thin-lens imaging theory, with the following results:
\begin{eqnarray}
| K^{(m)}_{1',2'}(\rhovec_{1}, \rhovec_{2})| &=& \Biggl |\frac{k_{0}M}{2\pi L_{R}} \! \int \! {\mathrm d}\rhovec'  e^{-i k_{0}(2 \rhovec_{2} \cdot \rhovec'- |\rhovec'|^2)/2L_R}  \Biggr. \nonumber \\[.12in] && \Biggl. \times\,\, K_{1,2}^{(m)}(M \rhovec_{1},\rhovec') T(\rhovec') \Biggr |, \label{corr:magnitude}
\end{eqnarray}
where $M \equiv - d_{2}/d_{1}$ is the signal-arm magnification factor.  For a sufficiently large bucket detector we can approximate the integrals in \eqref{GI:corrseparate} as covering the entire plane, viz., 
\begin{eqnarray}
\lefteqn{C_{m} \int_{\mathcal{A}_2}  d \rhovec_{2}\,   |K^{(m)}_{1',2'}(\rhovec_{1}, \rhovec_{2})|^{2}}
\nonumber \\[.12in] &\approx& C_{m} \int\!{\rm d} \rhovec_{2}\,   |K^{(m)}_{1',2'}(\rhovec_{1}, \rhovec_{2})|^{2} \\[.12in]  &=& M^{2} C_{m} \int\! {\rm d} \rhovec\,   |K^{(m)}_{1,2}(M \rhovec_{1}, \rhovec)|^{2} |T(\rhovec)|^2\,,
\end{eqnarray}
for $m = n, p$, where we the last equality follows from Parseval's theorem. In this limit, $C'(\rhovec_{1}) =  M^2 C(M \rhovec_1)$, where $C(\rhovec_{1})$ is given by \eqref{GI:corr}.  Hence choosing $d_{1} = d_{2} = 2f$  will yield an inverted version of the object-plane ghost image. Image resolution and field of view are then determined by the phase-sensitive and phase-insensitive coherence properties of the object-plane fields, and the placement of the detectors relative to this plane only determines the signal-arm optics that are needed to obtain this object-plane ghost image.

\section{Discussion}
\label{discussion}

The fundamental source property that enables acquisition of a ghost image---whether the source is classical or quantum---is the non-zero cross-covariance between the photon-flux densities of the two detected fields, i.e., the cross-correlation of the photon-flux densities minus the product of their mean values.  In particular, the product of mean values generates the background term, while the cross-covariance produces the image-bearing terms.  For zero-mean, Gaussian-state sources, the cross-correlation of the photon-flux densities, which is a fourth-order field moment, reduces to a sum of terms involving second-order field moments.  Consequently, both phase-sensitive and phase-insensitive field-operator cross-correlations can contribute to the ghost image. In the Appendix~\ref{appendix} we will show that any pair of phase-sensitive and phase-insensitive cross-correlation functions can be obtained with two classical Gaussian-state fields, so long as there are no restrictions on these fields' auto-correlation functions. In this respect, the ghost image does {\em not} contain any quantum signature {\em per se}. However, if we compare sources that have identical auto-correlation functions, we find that nonclassical fields with low brightness and maximum phase-sensitive cross correlation offer a spatial resolution advantage in the source's near field and a field-of-view expansion in its far field. The field of view in the near field and the resolution in the far field are determined by the beam sizes at the source, and hence are identical for classical and nonclassical fields. 

A number of recent publications have implied that ghost imaging with thermal-state light cannot be explained by classical electromagnetic theory in combination with semiclassical photodetection theory, but that a strictly quantum-mechanical interpretation involving nonlocality must be used to understand such experiments \cite{Scarcelli,Shih:QI}.  A key conclusion from our paper, however, is that the classical theory of ghost imaging is {\em quantitatively indistinguishable} from the quantum theory of ghost imaging for any optical source that is in a classical state, regardless of the propagation geometry.  Here, a classical state is one whose density operator has a proper $P$-representation, so that its photodetection statistics can be correctly quantified with classical, stochastic-field electromagnetism and detector shot noise. Thermal light--whether it is broadband, such as natural illumination, or narrowband, such as thermalized laser light---falls precisely within this category of Gaussian states.  Therefore experiments utilizing thermal light sources alone cannot validate the quantum description.  Furthermore, and perhaps more critically, there is no nonlocal interaction in thermal-light ghost imaging.  In particular, because the joint state of the signal and reference beams is classical---in the sense noted above---it \em cannot\/\rm\ lead to a violation of  the Clauser-Horne-Shimony-Holt (CHSH) inequality \cite{CHSH}.   We reiterate that it is the non-zero cross-covariance between the photon-flux densities of the signal and reference fields that is responsible for the image-bearing terms obtained from the Fig.~\ref{GI:propagation} setup. For Gaussian-state sources, this detection-plane cross-covariance is found, by moment factoring, from the phase-insensitive and phase-sensitive field cross-correlation functions. These detection-plane field correlations follow, in turn, from propagation of the corresponding source-plane field correlations through $L$\,m of free space.  Thus, two classical fields that are generated in a correlated fashion at a source, yet propagating paraxially in two different directions, will still exhibit spatio-temporal correlations on transverse planes that are equidistant from the source, even though these planes may be physically separated from each other. This  concept is both well known in and central to classical statistical optics \cite{Goodman:Stat,Mandel}.   It is not at all related to nonlocality in quantum mechanics, e.g., to violation of the CHSH inequality.

It is worth connecting some of the analysis presented in this paper with recent theory for the coherence properties of biphoton wave functions, which has led to an elegant duality between the partial entanglement of biphotons and the classical partial coherence of phase-insensitive fields \cite{Saleh:Duality}. As we have shown in Sec.~\ref{GaussianStates}, the biphoton state is the low-brightness, low-flux limit of the zero-mean jointly Gaussian state with zero phase-insensitive cross-correlation but maximum phase-sensitive cross-correlation. In this limit, the biphoton wave function {\em is} the phase-sensitive cross-correlation function between the signal and reference fields, and therefore the duality between phase-insensitive coherence propagation and the biphoton wave function propagation is rooted in the duality between phase-insensitive and phase-sensitive coherence propagation [cf. Eqs.~\eqref{FSProp:PIS} and \eqref{FSProp:PS}]. Furthermore, classical fields may also have phase-sensitive coherence.  Thus, to correctly understand the fundamental physics of quantum imaging, it is crucial to distinguish features that are due to the presence of this phase-sensitive correlation in the source fields from those that {\em require} this phase-sensitive correlation to be stronger than what is possible with classical (proper $P$-representation) states. The following examples clearly illustrate our point. When ghost imaging is performed with phase-sensitive light, image inversion occurs in the far field for both classical and quantum sources. This inversion is entirely due to the difference between the free-space propagation of phase-sensitive and phase-insensitive correlations, and it is not necessary for the phase-sensitive coherence to be stronger than classical. On the other hand, the background-free nature of ghost images formed with SPDC light arises from that source's phase-sensitive cross-correlation being much stronger the classical limit, as we showed in Sec.~\ref{contrast}.

In summary, we have used Gaussian-state analysis to establish a unified treatment of classical and quantum ghost imaging. Our analysis reveals that ghost-image formation is due to phase-sensitive and phase-insensitive cross-correlations between the signal and reference fields. Because arbitrary cross-correlations can be achieved by classical and quantum sources alike, image contrast is the only distinguishing feature between a source that is classical or quantum. In particular, we emphasize that a classical source with phase-sensitive cross-correlation can produce an {\em identical} image to that obtained with a biphoton source---up to a different contrast and hence signal-to-noise ratio---even for ghost-imaging configurations that utilize lenses, mirrors or other linear optical elements.
If we compare ghost images from classical and quantum sources having identical auto-correlations, thereby fixing the background level, the low-brightness quantum source offers resolution enhancement in near-field operation and field-of-view enhancement in far-field operation, in addition to higher contrast in both regimes.   Furthermore, because far-field spatial resolution and the near-field field of view are determined by source-plane beam size, they are identical for classical and quantum sources.  Finally, the conclusions in this paper are not contingent on having coincident object and detection planes.  They apply so long as the signal arm can be freely modified to transfer the object-plane correlations to the detection plane via an appropriately-positioned lens.

\section*{Acknowledgement}
This work was supported by the U. S. Army Research Office MURI Grant W911NF-05-1-0197.

\appendix*
\section{Classical Gaussian states with arbitrary cross correlations}
\label{appendix}

Let us use  $\hat{E}_{S}(\mathbf{x})$ and $\hat{E}_{R}(\mathbf{x})$ to denote the signal and reference field operators, where $\mathbf{x} = (\rhovec, t)$ conveniently combines their space and time arguments. In this appendix we will construct a zero-mean, jointly Gaussian, \em classical\/\rm\ state for these two quantum fields  that has arbitrarily prescribed phase-insensitive and phase-sensitive cross-correlation functions,
\begin{align}
K_{S,R}^{(n)}(\mathbf{x}_{1}, \mathbf{x}_{2}) & \equiv  \langle \hat{E}_{S}^{\dagger}(\mathbf{x}_{1}) \hat{E}_{R}(\mathbf{x}_{2}) \rangle\,, \label{CC:PIS} \\
K_{S,R}^{(p)}(\mathbf{x}_{1}, \mathbf{x}_{2}) & \equiv  \langle \hat{E}_{S}(\mathbf{x}_{1}) \hat{E}_{R}(\mathbf{x}_{2}) \rangle\,, \label{CC:PS}
\end{align}
respectively. We only require that both functions be sufficiently well behaved that they can be regarded as kernels which map the Hilbert space of square-integrable functions into itself.  Under this regularity condition we can  perform singular-value decompositions of these continuous kernels \cite{Deif} to obtain 
\begin{align}
K_{S,R}^{(n)}(\mathbf{x}_{1}, \mathbf{x}_{2}) &= \sum_{m=1}^{\infty} \eta_{m} \phi_{m}^{*}(\mathbf{x}_{1}) \Phi_{m} (\mathbf{x}_{2})\,,\\
K_{S,R}^{(p)}(\mathbf{x}_{1}, \mathbf{x}_{2}) &= \sum_{m=1}^{\infty} \mu_{m} \psi_{m}(\mathbf{x}_{1}) \Psi_{m}(\mathbf{x}_{2})\,,
\end{align}
where $\{\phi_{m}(\mathbf{x}_{1})\}$, $\{\Phi_{m}(\mathbf{x}_{1})\}$, $\{\psi_{m}(\mathbf{x}_{1})\}$ and $\{\Psi_{m}(\mathbf{x}_{1})\}$, for $1\le m <\infty$, are four complete and orthonormal sets spanning square-integrable functions, and the coefficients $\eta_{m}$ and $\mu_{m}$ are real, finite, and non-negative for all $m$.

Suppose we define two pairs of free-space, paraxial field operators, $\{\hat{E}_{S'}(\mathbf{x}), \hat{E}_{R'}(\mathbf{x})\}$ and $\{\hat{E}_{S''}(\mathbf{x}), \hat{E}_{R''}(\mathbf{x})\}$, having the modal expansions
\begin{align}
\hat{E}_{S'}(\mathbf{x}) &= \sum_{m=1}^{\infty} \hat{a}_{S'\!, m}  \Phi_{m}(\mathbf{x}) \\
\hat{E}_{R'}(\mathbf{x}) &= \sum_{m=1}^{\infty} \hat{a}_{R'\!, m}  \phi_{m}(\mathbf{x})\,,
\end{align}
and 
\begin{align}
\hat{E}_{S''}(\mathbf{x}) &= \sum_{m=1}^{\infty} \hat{a}_{S''\!, m}  \psi_{m}(\mathbf{x}) \\
\hat{E}_{R''}(\mathbf{x}) &= \sum_{m=1}^{\infty} \hat{a}_{R''\!, m}  \Psi_{m}(\mathbf{x}) \,.
\end{align}
In these expansions,  $\{\hat{a}_{\ell,m}\}$, for $\ell = S',S'',R',R''$ and $1 \le m <\infty$, is a set of photon annihilation operators, with the canonical commutation relations $[ \hat{a}_{\ell,m}, \hat{a}^{\dagger}_{\ell',m'} ] = \delta_{\ell,\ell'} \delta_{m,m'}$ and $[ \hat{a}_{\ell,m}, \hat{a}_{\ell',m'} ] = 0$.  

Now, let us put the modes associated with the $\{\hat{a}_{\ell,m}\}$ into a zero-mean, jointly Gaussian state whose only non-zero phase-insensitive cross-correlations are, 
\begin{equation}
\langle \hat{a}_{S', m}^{\dagger} \hat{a}_{R', m} \rangle = 2 \eta_{m}\,, \label{PIS:cc}
\end{equation}
and whose only non-zero phase-sensitive cross-correlations are
\begin{equation}
\langle \hat{a}_{S'', m} \hat{a}_{R'', m} \rangle = 2 \mu_{m} \,, \label{PS:cc}
\end{equation}
whence
\begin{align}
\langle \hat{E}_{S'}^{\dagger}(\mathbf{x}_{1}) \hat{E}_{R'}(\mathbf{x}_{2}) \rangle &= 2 K_{S,R}^{(n)}(\mathbf{x}_{1}, \mathbf{x}_{2}) \\
\langle \hat{E}_{S''}(\mathbf{x}_{1}) \hat{E}_{R''}(\mathbf{x}_{2}) \rangle &= 2 K_{S,R}^{(p)}(\mathbf{x}_{1}, \mathbf{x}_{2}).
\end{align}

Classical Gaussian states must have correlations that obey the Cauchy-Schwarz inequalities from \eqref{CS:PISq} and \eqref{CS:PSc}, see \cite{ShapiroSun}. Thus Eqs.~\eqref{PIS:cc} and \eqref{PS:cc} imply that the modal auto-correlations must obey
\begin{equation}
\langle \hat{a}_{S',m}^{\dagger} \hat{a}_{S',m} \rangle  \langle \hat{a}_{R', m}^{\dagger} \hat{a}_{R', m} \rangle \geq 4 \eta_{m}^{2}\,,
\end{equation}
and 
\begin{equation}
\langle \hat{a}_{S'',m}^{\dagger} \hat{a}_{S'',m} \rangle  \langle \hat{a}_{R'',m}^{\dagger} \hat{a}_{R'',m} \rangle \geq 4 \mu_{m}^{2}\,,
\end{equation}
for $1\le m <\infty$. We will take the modal auto-correlations to equal these lower bounds, by assuming that 
\begin{align}
\langle \hat{a}_{S',m}^{\dagger} \hat{a}_{S',m} \rangle &=  \langle \hat{a}_{R',m}^{\dagger} \hat{a}_{R',m} \rangle  = 2 \eta_{m} \label{auto1}\\
\langle \hat{a}_{S'',m}^{\dagger} \hat{a}_{S'',m} \rangle &=  \langle \hat{a}_{R'',m}^{\dagger} \hat{a}_{R'',m} \rangle  = 2 \mu_{m},
\label{auto2}
\end{align}
We shall also assume that all modal correlations---aside from those which have already been specified ---vanish.  Equations~\eqref{PIS:cc}, \eqref{PS:cc}, \eqref{auto1} and \eqref{auto2} then determine the zero-mean, jointly Gaussian state of the four fields, $\{\hat{E}_{S'}(\mathbf{x}), \hat{E}_{R'}(\mathbf{x}),\hat{E}_{S''}(\mathbf{x}), \hat{E}_{R''}(\mathbf{x})\}$, which is the tensor product of  zero-mean, jointly Gaussian states of  $\{ \hat{E}_{S'}(\mathbf{x}),\hat{E}_{R'}(\mathbf{x}) \}$ and $\{ \hat{E}_{S''}(\mathbf{x}),\hat{E}_{R''}(\mathbf{x}) \}$,  because all their cross-correlations are zero. 
Defining the signal and reference fields via
\begin{align}
\hat{E}_{S}(\mathbf{x}) &= (\hat{E}_{S'}(\mathbf{x}) + \hat{E}_{S''}(\mathbf{x}))/\sqrt{2}\\
\hat{E}_{R}(\mathbf{x}) & = (\hat{E}_{R'}(\mathbf{x}) + \hat{E}_{R''}(\mathbf{x}))/\sqrt{2}\,,
\end{align}
thus yields a pair of field operators that are in a zero-mean, jointly Gaussian, \em classical\/\rm\ state with the desired phase-sensitive and phase-insensitive cross-correlation functions.


\begin{thebibliography}{10}
\bibitem{GIvariations} Other ghost imaging configurations replace the scanning pinhole detector with a CCD array for parallel data acquisition, or separate the object plane and the detection plane to allow greater flexibility in implementation, or image in reflectance rather than transmission. These variations do not affect the fundamental physics that governs ghost imaging, and, with the exception of separating the object and detection planes, will not be discussed herein.
\bibitem{coherence_separability} The signal and reference fields obtained from spontaneous parametric downconversion (SPDC) will not have space-time correlation functions that take these specific forms, because of the space-time coupling that is inherent in SPDC phase-matching relations.  However, these assumptions, which are commonly-employed in coherence theory, simplify the analytical treatment without compromising the fundamental physics that yield a ghost image \cite{Mandel,Goodman:Stat}.
\bibitem{PIS:proofsketch} Because zero-mean Gaussian states are completely determined by their correlation functions,  \eqref{CS:PISq}  implies, via the moment-factoring theorem for Gaussian states, that all higher-order moments of the signal and reference fields will be admissible moments for a pair of classical stochastic processes. Therefore, the characteristic functional for this Gaussian state---which is a generating function for all of its moments---must be an admissible characteristic functional for a pair of classical stochastic processes. Then, via Sec.~\ref{semiclVSquantum}, the zero-mean jointly Gaussian state is representable as a classical statistical mixture of $\hat{E}_{S}$ and $\hat{E}_{R}$'s coherent states (whose sample functions constitute the eigenfunctions of the coherent states, and the probability density of their occurrence constitutes the mixture-weights of these coherent states), viz., the joint state has a proper $P$-representation.
\bibitem{autoVScross} The distinction between auto- and cross-correlation propagation is irrelevant here, because both the signal and reference beams undergo identical transformations. Thus, even though we state our results only for the cross-correlation functions, these results also apply to auto-correlation propagation.
\bibitem{FOV}Field of view usually refers to a solid-angle region, but we will use the intensity radius at a transverse plane as our field-of-view measure.
\end{thebibliography}
\end{document}